\documentclass[12pt]{article}
\usepackage{epsfig,latexsym,amssymb,amsmath}

\def\be{\begin{equation}}
\def\beq{\begin{equation}}
\def\eeq{\end{equation}}
\newcommand{\en}{\end{equation}}
\def\ba{\begin{eqnarray}}
\def\bea{\begin{eqnarray}}
\def\ea{\end{eqnarray}}
\def\eea{\end{eqnarray}}
\newcommand{\eqa}{\begin{eqnarray}}
\newcommand{\ena}{\end{eqnarray}}

\def\Tr{{\rm Tr}}

\renewcommand{\l}{\lambda}

\renewcommand{\b}{\beta}
\renewcommand{\a}{\alpha}
\newcommand{\n}{\nu}
\newcommand{\m}{\mu}

\newcommand{\ep}{\varepsilon}

\renewcommand{\d}{\delta}

\renewcommand{\r}{\rho}
\newcommand{\s}{\sigma}

\newcommand{\prt}{\partial}

\newcommand{\no}{\nonumber}
\newcommand{\non}{\nonumber \\}

\begin{document}

\begin{titlepage}
\vskip0.5cm
\begin{flushright}
DFTT 18/05\\
\end{flushright}
\vskip 2.4cm
\begin{center}
{\Large\bf  A new large N phase transition in YM$_\mathbf{2}$}
\end{center}
\vskip 1.cm \centerline{ A. Apolloni, S. Arianos and A. D'Adda}
\vskip0.6cm \centerline{\sl  Dipartimento di
Fisica Teorica dell'Universit\`a di Torino and} \centerline{\sl
Istituto Nazionale di Fisica Nucleare, Sezione di Torino}
\centerline{\sl via P.Giuria 1, I-10125 Torino, Italy
\footnote{e--mail: {\tt
apolloni, arianos, dadda@to.infn.it}}} \vskip 2cm
\begin{abstract}
Inspired by the interpretation of two dimensional Yang-Mills theory on a cylinder
as a random walk on the gauge group, we point out the existence of a large $N$
transition which is the gauge theory analogue of the cutoff transition in random
walks. The transition occurs in the strong coupling region, with the 't Hooft 
coupling scaling as $\alpha \log{N}$, at a critical value of $\alpha$ ($\alpha =4$
on the sphere). The two phases below and above the transition are studied in
detail. The effective number of degrees of freedom and the free energy are found
to be proportional to $N^{2-\frac{\a}{2}}$  below the transition and to vanish
altogether above it. The expectation value of a Wilson loop is calculated to 
the leading order and found to coincide in both phases with the 
strong coupling value.
  
\end{abstract}
\end{titlepage}

\section*{Introduction}   

Yang-Mills theory in two dimensions is at the intersection of many different
fields of theoretical physics. It is one example of non trivial completely 
solvable gauge theory~\cite{Rusakov:1990rs,Migdal:1975zg}, in which  both perturbative 
and non perturbative effects can be studied. Its large $N$ expansion
has been proved to describe a two dimensional string theory
~\cite{Gross:1993hu,Gross:1993yt}, namely a theory of
branched coverings on a two dimensional Riemann surface. Non trivial topological
sectors in the unitary gauge also seem to be related to matrix string 
states~\cite{Billo:1998fb}.
Its partition function on a torus can be described in terms of a gas of free
fermions~\cite{Douglas:1993xv,Minahan:1993np,Minahan:1993mv,Caselle:1993gc}, 
and the kernel on a cylinder by the evolution of a system of $N$ free
fermions on a circle, namely by the Sutherland model.

Some new connections between two dimensional gauge theories and statistical 
mechanical systems were pointed out in \cite{D'Adda:2001nf} where two 
dimensional gauge theories of the symmetric group $S_n$ in the large $n$ limit 
were investigated.
Gauge theories of $S_n$ also describe $n$-coverings of a Riemann surface and 
hence they are closely related to two dimensional Yang-Mills theories; the 
relation being essentially provided by Frobenius formula that relates 
U($N$) characters (in a representation with $n$ boxes in the Young diagram) to 
the corresponding characters of the $S_n$ group.

It was shown in \cite{D'Adda:2001nf} that the partition function of a gauge
theory on a disc or a cylinder can be interpreted in terms of random walks on 
the gauge group, whose initial and final positions are the holonomies at the ends of the
cylinder\footnote{In a disc the starting point is the identity of the group, and
on a sphere both starting and ending points are the identity.} and the number 
of steps is the area of the surface.
Although the focus in \cite{D'Adda:2001nf} was on the discrete group $S_n$ the 
argument can be trivially extended, as it is shown in the Appendix of the 
present paper, to continous Lie groups. 
Similar results were independently obtained in \cite{deHaro:2004id}.
It is well known in random walks theory that after a certain number of steps the
end point of the walk becomes independent of the starting point: the walker has 
lost any memory of the point he started from. The critical number of steps after
which that happens can be exactly calculated in a number of situations, and the
corresponding transition is known as  cutoff transition.
Given the correspondence between random walks on the group and gauge theory 
on a cylinder, one expects to find the cutoff transition also in gauge theories. 
Indeed it was found in \cite{D'Adda:2001nf} that for an $S_n$ gauge theory 
where the holonomy on each elementary plaquette is given by a single
transposition\footnote{In terms of the string interpretation this means that in
each plaquette there is a single quadratic branch point connecting two of the
$n$ sheets of the world sheet.}, a cutoff transition occurs in 
the large $n$ limit when the number of plaquettes (and hence the area) is 
$\frac{1}{2} n \log{n}$, in agreement with previous results in random 
walks \cite{Diaconis:1981}.
Models with more general Boltzmann weight for the plaquettes have a richer
structure of phase diagrams~\cite{D'Adda:2001nf}.
The stringy interpretation of the cutoff transition in the $S_n$ gauge theory is 
the following: beyond the transition the string world sheet is completely connected
in the large $n$ limit, while before the transition the world sheet consists
of a large connected part and of  a small fraction (in fact vanishing
in the large $n$ limit) of disconnected parts.

Another well known correspondence relates random walks with random graphs \cite{Pak:2001}, 
that is  graphs obtained by randomly connecting $n$ points with $p$ links. 
These can be put in correspondence with random walks on $S_n$ made of $p$ steps,
each step consisting of a simple transposition.
Two types of transitions are known in the large $n$ large $p$ double scaling 
limit in random graphs: a percolation transition at a critical value 
$\beta = \beta_c$ when $p = \beta n$ and the cutoff transition at 
$\alpha = 1/2$ when $p = \alpha n \log{n}$. Beyond the transition, namely for 
$p > 1/2 n \log{n}$  all the $n$ points are connected whereas before the
transition a vanishing fraction of disconnected points survive.

It is rather natural at this point to look for a similar transition in the large
$N$ limit of U($N$) gauge theories.
A large $N$ phase transition on a sphere and on a cylinder in two dimensional 
Yang-Mills theories - the Douglas-Kazakov phase transition - has been known for
quite some time.
However this is not a cutoff transition. In a cutoff transition the partition 
function on a disc for instance becomes independent on the holonomy on the 
border of the disc, and this
is not the case in the Douglas-Kazakov transition. Besides the  Douglas-Kazakov transition
occurs at a finite value of the 't Hooft coupling whereas from the previous
examples it appears that the cutoff 
transition occurs when the area  scales as  $\log{N}$ at large $N$.
From this point of view the Douglas-Kazakov transition appears more similar to 
the percolation transition in random graphs, although a precise correspondence 
is still to be found.

The existence of the cutoff transition in the large $N$ limit of $2$D 
Yang-Mills on the sphere when the 't Hooft coupling scales as $\log{N}$ was
proved by a simple argument in \cite{D'Adda:2001nf}.
The present paper is devoted to  study such transition further on both the sphere
and the cylinder, in order to characterize its phases and give some physical
interpretation.
The paper is organized as follows: in Section 1 we review the large $N$
Douglas-Kazakov transition and its physical interpretation. In Section 2
we introduce the cutoff transition on the sphere and study the phases above and
below the transition. Section 3 is devoted to the transition on the disc and on
the cylinder and Section 4 to the calculation of the expectation value of Wilson
loops.


\section{Large $N$ transition}

The partition function of a pure gauge theory on an arbitrary orientable 
two-dimensional manifold $\mathcal{M}$ of genus $G$, $p$ boundaries and area
 $\tilde{A}$ has been known for many years \cite{Rusakov:1990rs,Migdal:1975zg}:
\bea 
Z_{\mathcal{M}} & = & \int[\mathcal{D}A^{\m}]e^{-\frac{1}{4\l^2}\int_{\mathcal{M}}d^2x
\sqrt{g}\Tr(F^{\m\n}F_{\m\n})} \non & = & \sum_{r}\chi_r(g_1)\cdots\chi_r(g_p)
d^{2-2G-p}_re^{-\frac{A}{2N}C_2(r)}\;.
\label{first} 
\eea
The sum runs over all irreducible representations of the gauge group, $\l$ is the gauge 
coupling, $\chi_r(g_i)$ is the character of the holonomy $g_i$ in the representation $r$ 
and $C_2(r)$ is the quadratic Casimir operator in the representation $r$. $A$ is related 
to the actual area of $\mathcal{M}$ through $A=\l^2N\tilde{A}$.

We  consider $G=0$ manifolds with at most 2 boundaries; i.e. spheres, discs and cylinders. Moreover, we will confine our analysis to the unitary groups $U(N)$ and $SU(N)$.
A third order phase transition in the large $N$ limit was discovered in the case of a 
sphere by Douglas and Kazakov \cite{Douglas:1993ii} at a critical value $A=\pi^2$ of the
rescaled area $A$. This transition appears to separate a weak coupling ($A<\pi^2$) from a 
strong coupling ($A>\pi^2$) regime. 
These results were generalized to the case of a cylinder in \cite{Caselle:1993mq,
Gross:1994mr} where the phase transition was also interpreted as a result of instanton
condensation.
 
The partition function on the sphere can be written as a sum over the set of 
integers $n_1>n_2>\cdots>n_N$ that label the irreducible representations of 
SU($N$) and U($N$)\footnote{In the case of U($N$) the extra condition  $n_N\ge-\frac{N-1}{2}$
must be imposed.}:

\beq \label{chp}
Z=e^{-\frac{A}{24}(N^2-1)}\sum_{n_1>n_2\cdots>n_N}\prod_{i<j}(n_i -n_j)^2 
~~e^{-\frac{A}{2N}\sum_{i=1}^{N}(n_i)^2}\;,
\eeq

\begin{figure}[ht]
\begin{minipage}[t]{0.45\textwidth}
\begin{center}
\includegraphics[width=5.0 cm]{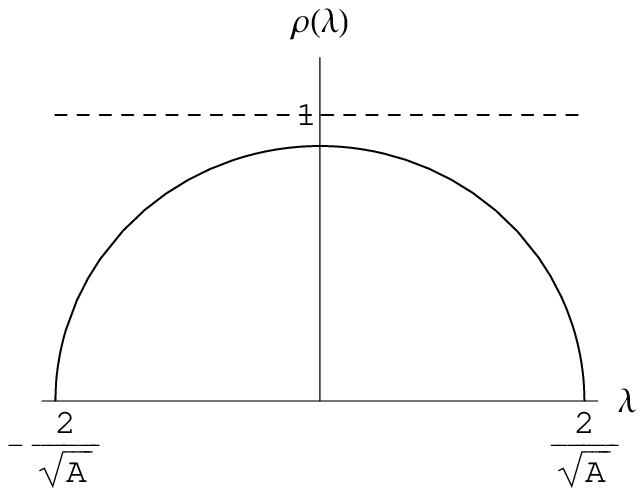}
\end{center}
\caption{\small{Eigenvalue  distribution for $A \leq \pi^2$ }}
\label{primaDk}
\end{minipage}
\hfill
\begin{minipage}[t]{0.45\textwidth}
\begin{center}
\includegraphics[width=5.0 cm]{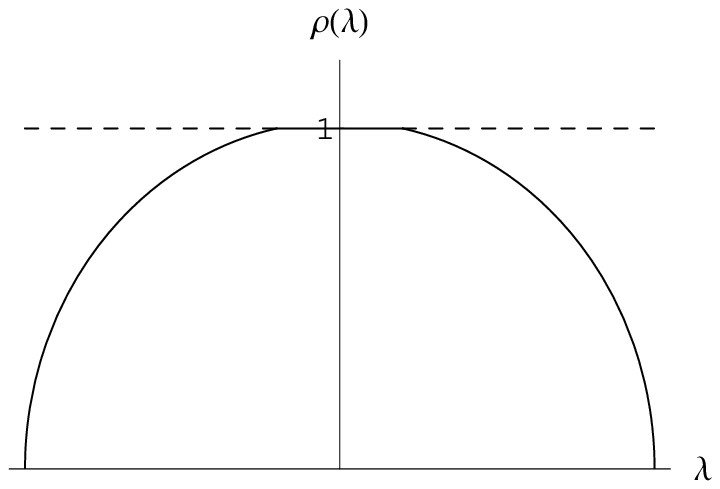}
\end{center}
\caption{\small{Eigenvalue distribution for $A > \pi^2$ (i.e. DK
transition)}} \label{dopoDk}
\end{minipage}
\end{figure}

The existence of a phase transition at the critical value $A=\pi^2$ can be
easily derived from (\ref{chp}) by noticing that the partition function 
(\ref{chp}) is exactly the same as the one of a gaussian hermitian matrix model
 but with the integral over the eigenvalues replaced by the discrete sum over
 the integers $n_i$.
The solution of a gaussian hermitian matrix model in the large $N$ limit is
given by Wigner's semicircle distribution law for the eigenvalues:
\beq
|\r(\lambda)|=\frac{A}{2\pi}\sqrt{\frac{4}{A}-\lambda^2}\;.
\label{wigner}
\eeq
where the continuum variables 
$$x=\frac{i}{N}\;,\quad \lambda(x)=\frac{n_i}{N}\;.$$
and the corresponding density of eigenvalues
\beq
\r(\lambda)=\frac{\prt x}{\prt \lambda}
\eeq
have been introduced.
In (\ref{chp}) however the would-be eigenvalues $n_i$ are distinct integers and as
a consequence the corresponding density $\r(\lambda)$ in the large $N$ limit is 
constrained by:
\beq \label{tr}
|\r(\lambda)|\le 1\quad \forall\;\lambda\;.
\eeq
Hence the Wigner semicircle solution is acceptable only in the weak coupling
phase, namely for $A \le \pi^2$ where the condition (\ref{tr}) is fulfilled.
In fact the maximum of $|\r(\lambda)|$ occurs at $\lambda=0$, it increases with the area
$A$ and becomes equal to $1$ at $A=\pi^2$, as easily seen from (\ref{wigner}). 
The solution in the strong coupling phase $A > \pi^2$ was found in 
\cite{Douglas:1993ii} and is expressed in terms of elliptic integrals.
In this phase a finite fraction of the eigenvalues condenses, namely the
distribution $\r(\lambda)$ is flat and equal to one in a symmetric interval
around $\lambda=0$ as shown in fig. \ref{dopoDk}.

\subsection{Configuration space}

It is well known \cite{Douglas:1993xv,Minahan:1993np,Minahan:1993mv,Caselle:1993gc}
that the partition function of two dimensional Yang-Mills
theories with gauge group U($N$) can be interpreted in terms of a gas of $N$ free fermions
on a circle described by a Sutherland-Calogero model. In particular, if we denote by 
${\cal K}_2 (\theta,\phi; A)$ the kernel on a 
cylinder of scaled area $A$ and with the U($N$) holonomies at the two ends given by the 
invariant angles $\theta_i$ and $\phi_i$, it was shown that ${\cal K}_2 (\theta,\phi; A)$ 
can be interpreted as the propagator in a time $A$ from an initial configuration where the 
positions of the $N$ fermions on the circle are given by $\theta_i$ and a final
configuration with positions labeled by $\phi_i$.
The partition function on the sphere of area $A$ is a particular case where the initial
and final configurations are just $\theta_i=\phi_i=0$ for all $i$'s, namely the amplitude 
for a process
where all fermions start at the origin and come back to the origin after a time $A$.
By a modular transformation on the kernel of the cylinder one finds that the  integers 
$n_i$ labeling the irreducible representations of U($N$) are just the discrete momenta of
the fermions on the circle. 
While in the momentum representation the Douglas-Kazakov phase transition can be
interpreted as fermion condensation, in the configuration representation it can be seen in
terms of instantons condensation \cite{Caselle:1993mq,Gross:1994mr}.
In fact, while going from the initial $\theta_i=0$ configuration to the final $\phi_i=0$
configuration, a fermion can in principle wind an arbitrary number of times around the
circle. These winding (instantons) configurations do not contribute in the large $N$ limit
to the weak coupling phase, as shown by the following simple argument.
Consider the Wigner distribution (\ref{wigner}) of momenta in the weak coupling phase. 
The maximum allowed momentum is $n_{\rm max}= \frac{2}{\sqrt{A}}$, hence the maximum shift
in position for a single fermion in the time $A$ is given by
\beq
\Delta \theta_{\rm max} = A \frac{2}{\sqrt{A}}= 2 \sqrt{A}
\label{deltatheta}
\eeq
The existence of winding trajectories requires this shift in position to be at least $2 \pi$,
namely $A$ to be greater of $\pi^2$. 
Hence the critical value of $A$, where the Douglas-Kazakov phase transition occurs, marks
the point where instantons condense, and contribute to the functional integral in the
large $N$ limit.

A more detailed understanding of the Douglas-Kazakov phase transition can be
achieved by introducing in the large $N$ limit the density
$\rho (\theta,t)$ of  fermions in the position $\theta$ at a given time 
(=area) $t$. Due to the compact nature of the configuration space $\rho 
(\theta,t)$ is defined in the interval $-\pi \leq \theta \leq \pi$ with
$\rho (-\pi,t)=\rho(\pi,t)$. 
Matytsin proved \cite{Matytsin:1993iq} that if the evolution equation of the 
fermions is given by the Calogero-Sutherland model then the density 
$\rho (\theta,t)$ is governed by the Das-Jevicki equation \cite{Das:1990ka}
which admits Wigner semicircular distribution of radius $r(t)$ as a solution:
\beq
 \rho(\theta,t)=\frac{2}{\pi r(t)^2}\sqrt{r(t)^2-\theta^2}\;\;,\quad\; |\theta| \leq r(t)
 \label{kappa2}
 \eeq
 with $r(t)$ satisfying the differential equation $\frac{d^2 r(t)}{dt^2} +
 \frac{4}{r(t)^3}=0$. On a sphere of area $A$ the boundary conditions are $
 r(0)=r(A)=0$ and the differential equation has the solution:
 \beq
 r(t)= 2 \sqrt{\frac{t (A-t)}{A}}
 \label{radius}
 \eeq
 The solution given by (\ref{kappa2}) and (\ref{radius}) is valid provided the
 support of the density function $\rho(\theta,t)$ is in the interval
 $[-\pi,\pi]$ at all $t$, namely provided $r(t) \leq \pi$. The maximum value for
 the radius $r(t)$ occurs for $t=A/2$ and is $r_{max} = \sqrt{A}$. Hence the
 condition for the validity of the Wigner semicircular solution is $A \leq
 \pi^2$. Beyond the critical value $A=\pi^2$ the fermions "realize" that the
 space they live in is compact, instantons effects become important and (\ref{kappa2})
 is not an acceptable solution any longer.
 On the cylinder a similar phase transition occurs in general, at a critical value
 of the area that depends on the holonomies at the boundaries\footnote{However
 with particular conditions at the boundary the phase transition may also be
 absent, see for instance \cite{Zelditch:2003mt}.}.
 If the distribution of the invariant angles of the holonomies at the boundaries is the
 Wigner semicircular distribution, then the critical area can be calculated exactly in the
 same way as for the sphere \cite{Caselle:1993mq}. For a more general discussion see 
 \cite{Gross:1994mr}. 


\section{Large $A(N)$ phase transition} \label{lA}

\begin{figure}[ht]
\begin{center}
\includegraphics[width=8.5 cm]{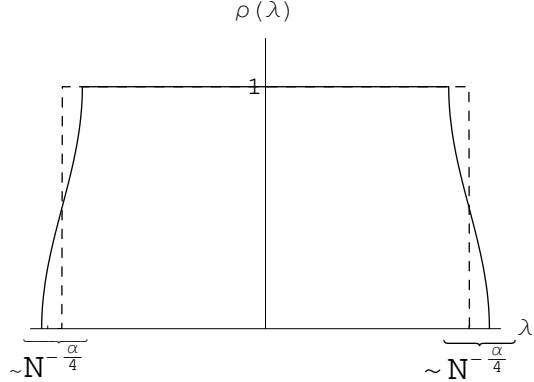}
\end{center}
\caption{\small{Eigenvalues distribution for $0<\a<4 $
}} \label{prima}
\end{figure}

It is apparent from the discussion in the previous section and from the explicit form 
(\ref{chp}) of the partition function that as the area of the sphere
increases the distribution of the "momenta" $n_i$ (i.e. of the integers that label the 
U($N$) representations)
becomes more and more similar to a double step function, like the one drawn with dashed lines in fig.
\ref{prima}. 
In fact for very large areas the attractive quadratic potential tends to dominate over the
repulsive force produced by the Vandermonde determinant. The double step distribution  
corresponds to the trivial representation of U($N$)
in which all characters are identical irrespective of their argument. If this distribution
dominates the functional integral then the kernel on a disc or on a cylinder becomes
independent from the holonomies at the boundaries.

As already mentioned in the introduction and discussed in the Appendix, the partition 
functions on the disc and on the cylinder may be interpreted in terms of random walks
on the group manifold.  From this point of view the very large area phase, where the sum
over the irreducible representations is dominated by the trivial representation corresponds
to a walk which is so long that the walker has lost any notion of the starting point. 
The transition where this situation sets in is known in random walk theory  as  
"cutoff transition", and the same term will be used here.

The cutoff regime occurs for areas larger than a critical $N$ dependent value $A_c(N)$
which was found in \cite{D'Adda:2001nf}. The argument is very simple: consider the trivial
double step representation $R_0$ that minimizes the Casimir term $\sum_i n_i^2$
 \beq 
R_0 :~~~~\left\{n_1,\cdots,n_N\right\}= \left\{\frac{N-1}{2},\cdots,-\frac{N-1}{2} \right \}\;.
\label{cutoff}
\eeq
and compare its contribution to the partition function (\ref{chp}) with the one coming
from a representation $R_1$ in which $n_1$ has been increased by $1$, namely in which $
n_1=\frac{N-1}{2} + 1$. In other words we look for the value of $A$ at which $R_0$ ceases
to be dominant. 
A simple calculation shows that the ratio between the two contributions is
\begin{equation}
\frac{Z_0 (A,N,R_0)}{Z_0(A,N,R_1)}=\frac{e^\frac{A}{2}}{N^2}.
\label{comparison}
\end{equation}
This ratio is larger than $1$ (hence $R_0$ dominates and we are in the cutoff phase)
if $A > 4 \log N$.

In order to study this phase transition in more detail, it is convenient  to parametrize 
the area $A$ by rescaling it with $\log N$:
\beq
A = \alpha \log N + \beta
\label{param}
\eeq
From the previous argument we expect the cutoff transition to occur at the critical value
$\alpha_c = 4$, separating two distinct phases .
So Yang Mills theory on a sphere seems to have four phases altogether: the first two at 
$\alpha=0$ separated by the Douglas-Kazakov phase transition at $\beta=\pi^2$, the other 
two when the area is logarithmically rescaled with $N$ that are separated by the cutoff 
transition.

The aim of this section is to find the saddle point configuration and the free energy in 
the two phases above and below the cutoff point $\alpha_c = 4$: the cutoff phase $\alpha >
\alpha_c$ has been discussed above and it is  rather trivial, but the phase below
$\alpha_c$ appears as an interesting intermediate phase between the strong coupling phase in 
the Douglas-Kazakov transition and the cutoff phase. Hence we shall concentrate on this
in the rest of the section.    
\noindent

Let us consider again the partition function (\ref{chp}) and the corresponding action

\begin{equation}
S=2\sum_{i>j=1}^N\log|n_i-n_j|-\frac{A}{2N}\sum_{i=1}^Nn_i^2
\label{ac}
\end{equation}

We want to find the extremum of this action in the large $N$ limit, when $A$ is
parametrized as in eq. (\ref{param}). Since we expect  the saddle point distribution of
the "momenta" $n_i$ to be symmetric with respect to the origin ( $n_i \to -n_i$ ) we
shall perform the variation only with respect to symmetric configurations, that is we set

\beq
-M\le i\le M\,,\quad \textrm{with}\quad M=\frac{N-1}{2} \quad \textrm{and} \quad
n_{-i}=-n_i,
\label{not}
\eeq

Using this symmetry one can restrict the sums to non negative values of $i$
($i=1,\ldots,M$) and write the action as:

\begin{equation}
S=2 \sum_{i>j\geq 1}^M \log{(n_i-n_j)^2}+\sum_{i>j\geq 1}^M \log{(n_i+n_j)^2}-
\frac{A}{2M} \sum_{i=1}^M n^2_i \label{action}
\end{equation}

While in the cutoff phase $n_i=i$ for $i=1,\cdots,M$ below the cutoff transition we expect
a configuration of the type described in fig. \ref{prima}, namely:

\bea
n_i=i &  & i=1,\cdots, M-l \non
n_{M-l+\alpha}=M-l+\alpha+r_\alpha &  & \a=1,\cdots, l   \label{rpr}
\eea

where the  value of $l$ and the spectrum of the integers $r_{\alpha}$ are to be
determined. With these notations the action can be written as:

\begin{eqnarray}
S-S_0 & = & 4\sum_{\alpha=1}^l\sum_{j=1}^{M-l}\log{\bigg[\Big(1+\frac{r_\alpha}
{\alpha+j}\Big)\Big(1+\frac{r_\alpha}{M-l+\alpha+j}\Big)\bigg]}\non
& + & 2 \sum_{\alpha\neq \beta}\big(\log{\big( 1+\frac{r_\alpha+r_\beta}
{\alpha+\beta+2(M-l)} \big)}+\log{\big( 1+\frac{r_\alpha -r_\beta}
{\alpha-\beta}\big)}\big)\non
& - & \frac{A}{2M}\sum_{\alpha=1}^l (2(M-l+\alpha)r_\alpha+r_\alpha^2), 
\label{action2}
\end{eqnarray}

where $S_0$ represents the value of the action in the trivial representation $R_0$.
We shall assume that as $M \to \infty$ also $l \to \infty$ but at a slower rate
than $M$, namely $l/M \to 0$. We shall also assume that $l$ and $r_{\alpha}$ will
be of the same order in the large $M$ limit. These assumptions will be justified
{\it a posteriori}, in the sense that they will provide a stable saddle point in
the large $M$ limit when the area is scaled like $\log M$.  They are also
very reasonable assumptions: $l$ is of order $N$ in the strong coupling phase following the
Douglas-Kazakov transition and one expects that with the logarithmic rescaling
of the area it will shrink further by some power of $N$.

The first step in dealing with (\ref{action2}) is to make the dependence from $M$
explicit. By using the identity
\beq 
\prod_{j=1}^{M-l}(1+\frac{r_\alpha}{j+\alpha})(1+\frac{r_\alpha}{M-l+j+\alpha})
=\frac{(2M-2l+\alpha+r_\alpha)!}{(2M-2l+\alpha)!}\frac{\alpha !}{(\alpha 
+r_\alpha)!}
\label{prod}
\eeq
we can rewrite the action as

\begin{eqnarray}
S-S_0 & = & 4\sum_{\alpha=1}^l\log{\frac{(2M-2l+\alpha+r_\alpha)!}{(2M-2l+\alpha)!}\frac{\alpha !}{(\alpha 
+r_\alpha)!}}\non
& + & 2 \sum_{\alpha\neq \beta}(\log{(1+\frac{r_\alpha+r_\beta}{\alpha+\beta+2(M-l)})}+\log{(1+\frac{r_\alpha -r_\beta}{\alpha-\beta})})\non
& - & \frac{A}{2M}\sum_{\alpha=1}^l (2(M-l+\alpha)r_\alpha+r_\alpha^2), 
\label{action3a}
\end{eqnarray}

All  ratios of factorials in (\ref{action3a}) can be reduced to the form
$\log{\frac{(N+C)!}{N!}}$ with $\displaystyle N\rightarrow\infty$, 
$\displaystyle C \rightarrow \infty$ and $\displaystyle \frac{C}{N} 
\rightarrow 0 $. By repeated use of Stirling formula one finds, up to terms that
vanish as $\displaystyle \frac{C}{N} \rightarrow 0 $:
\beq
\log{\frac{(N+C)!}{N!}} \sim C \left[ \log N + f(\frac{C}{N}) \right]
\label{Stir}
\eeq
where
\beq
f(\frac{C}{N})=\log{(1+\frac{C}{N})}+\frac{\log{(1+\frac{C}{N})}}{\frac{C}{N}}-
1=\sum_{k=1}^{M-l}(-1)^{k-1} \frac{z^k}{k(k+1)}\;. 
\label{asyfunct}
\eeq
By using this asymptotic behaviour, and introducing  continuum variables in the
large $N$ limit, namely

\beq
 x=\frac{\alpha}{l} \quad r(x)=\frac{r_\alpha}{l} \quad \sum_\a= l\int dx\;.
 \label{contvar}
 \eeq

the action finally takes the form 
\begin{eqnarray} 
S-S_0 & = & 4l^2\int_0^1 dx \big[r(x)(\log{(\frac{M}{l})}-\log{(x+r(x))}+1+
\log{2})+x(\log{x} \non 
& - & \log (x+ r(x)))\big]+2 l^2 \int_0^1 dx \int_0^1\,
dy\log{(1+\frac{r(x)-r(y)}{x-y})}\non
& -& A l^2\int_0^1 dx r(x)    
\label{action3}
\end{eqnarray}
where  subleading terms (by powers of $\frac{l}{M}$ ) have been neglected.

Let us now parametrize the area $A$ according to eq. (\ref{param}) and write the
action as:

\begin{equation}
S-S_0=4 l^2 [F_0 \log{M}-F_1\log{l}+F_2] 
\label{action4}
\end{equation}
where  $F_0$, $F_1$ and $F_2$ are of order $1$ in the large $M$ and $l$ limit
and are given by:
\bea
F_0 & = & (1-\frac{\alpha}{4})\int_0^1 dx r(x) \non
F_1 & = & \int_0^1 dx r(x) \non
F_2 & = & \int_0^1 dx \Big[r(x)(-\log{(x+r(x))}+1-\log{2}-\frac{\beta}{4})+x(\log{x}-\log{(x+r(x))})\Big]\non
& + & \frac{1}{2} \int_0^1 dx \int_0^1 dy \log{(1+\frac{r(x)-r(y)}{x-y})} \no
\eea
In order to find the configuration that maximizes the functional integral in the
large $M$ limit we take the variation of (\ref{action4}) with respect to both
$l$ and $r(x)$.
The variation with respect to $l$ gives the equation:

\beq
(1-\frac{\alpha}{4}) \log{M}+ (\frac{F_2}{F_1}-\frac{1}{2})  =  \log{l} 
\label{maxl} 
\eeq
which shows that $l$ grows like $M^{1-\frac{\alpha}{4}}$. This is consistent
with what we expected: for $\alpha \to 0$ it gives $l \sim M$ as in the strong
phase beyond the Douglas-Kazakov transition, and at the cutoff point $\alpha=4$
the power vanishes as expected.

The variation with respect to $r(x)$ gives on the other hand
\beq
\int_0^1\,dy\frac{1}{x+r(x)-y-r(y)}-\log{(x+r(x))}  =  C 
\label{maxr} 
\eeq
with
\beq 
C=\log{l} - (1-\frac{\alpha}{4}) \log{M}+\frac{\beta}{4}+\log{2}=
\frac{F_2}{F_1}-\frac{1}{2}+\frac{\beta}{4}+\log{2}
\label{cc} 
\eeq
If one introduces  the new variable $\xi=x+r(x)$ and the  density function 
$\rho(\xi)=\frac{d x}{d \xi}$ with support in the interval $[0,a]$
 equation (\ref{maxr}) becomes\footnote{The lowest extreme of the interval
 is $r(0)$ which is zero by construction, the upper end $a$ is, according with the
 definition of $\xi$, $a=1+r(1)$.}:
\beq
\int_{0}^{a} d\eta \frac{\rho(\eta)}{\xi-\eta}-\log{\xi}=C\;.
\label{rhoeq}
\eeq

This is a standard type of equation for the density of eigenvalues in the large
$N$ limit of matrix models and  can be solved by standard analytic methods
(for a detailed discussion of this equation see for instance \cite{Kostov:1997bn}).
The resolvent function, whose discontinuity across the cut gives the density $\rho(\xi)$,
is given by:
\beq
H(\xi)=\log{\xi}+C-2\log\Big(\frac{\sqrt{\xi-a}+\sqrt{\xi}}{\sqrt{a}}\Big)
\label{resol}
\eeq
with the additional condition that for large $\xi$
\beq
 H(\xi) = \frac{1}{\xi} + {\cal O}\big(\frac{1}{\xi^2} \big)
\label{asb}
\eeq
The corresponding density is given by
\begin{equation} 
\rho(\xi)=\frac{2}{\pi}\arccos{(\sqrt{\frac{\xi}{a}})}\;.
\label{ro}
\end{equation}
This solution obviously describes, through  the symmetry
(\ref{not}), both the positive and the negative region of $n_i$.

The asymptotic condition (\ref{asb}) gives the two extra equations
\bea
a & = & 2 \;, \non
C & = & \log{2} \;. 
\label{bcond}
\eea
Eq. (\ref{ro}), together with the first of (\ref{bcond}) define $r(x)$ completely,
although in an implicit way. Hence all integrals involved in the definition of $F_1$
and $F_2$ can be calculated. The calculation can actually be done analitically and gives:
\beq
F_1 = \frac{1}{4}\;\;,~~~~~~~\frac{F_2}{F_1}= \frac{1}{2}-\frac{\beta}{4}
\label{condF}
\eeq
The second equation could have been derived independently from (\ref{cc}) and
(\ref{bcond}), so it constitutes a non trivial consistency check.
We can now write explicitely $l$ and the free energy $F$ in terms of the area $A(M)$:
\begin{eqnarray} 
l & = & e^{-\frac{\beta}{4}}M^{1-\frac{\alpha}{4}} = M e^{-\frac{A(M)}{4}}\nonumber\\ 
F & = & S-S_0=\frac{1}{2} l^2 =\frac{1}{2}e^{-\frac{\beta}{2}} M^{2-\frac{\alpha}{2}}=
\frac{1}{2} M^2 e^{-\frac{A(M)}{2} }
\label{fe}
\end{eqnarray}
Beyond the cutoff transition we have instead $F=l=0$, as the dominant eigenvalue distribution is given by the double step function sketched in fig. \ref{prima}.

The interpretation of these results from the point of view of the free fermion description
is very clear: beyond the cutoff  ($\alpha>4$) we are effectively in a zero temperature
situation where all fermions fill the Fermi sea with no holes. Below the cutoff 
instead ($\alpha < 4$) some excited fermions and the corresponding holes are present in 
proximity of the surface of the Fermi sea both on the positive and negative momentum side. 
The number of fermions above the sea level is given by $l$ in (\ref{fe}).
The ratio $\frac{l}{M}$  vanishes like $M^{-\frac{\alpha}{4}}$ in the
large $M$ limit. This distinguishes this phase from the strong coupling phase of the
Douglas-Kazakov transition, where such ratio remains finite , namely the number of fermions
above the Fermi sea level is of order  $M$. 

In spite of being described in terms of $N$ free fermions, the free energy is
proportional (with the standard 't Hooft scaling) to $N^2$, which reflects the original
number of degrees of freedom in a unitary $N \times N $ matrix model\footnote{The
description in the terms of $N$ fermions follows the integration over the angular
variables that reduces the matrix model to an integral over the eigenvalues.
 The ensueing Vandermonde determinant makes  the wave function describing the eigenvalues
 antisymmetric. As a consequence  the total momentum of $M$ left moving fermions is of order
$M^2$ rather than $M$, that is of the same order as the original number of bosonic degrees
of freedom.}. So it is not surprising that the free energy becomes proportional to
$l^2$ in presence of $l$ effective fermionic degrees of freedom when the area is rescaled
by a  $\log{M}$ factor. It is  as if the effective size of the original matrix had shrunk to $l
\times l$. A full understanding of the reduction of number of degres of freedom from the
point of view of the original gauge degrees of freedom is still wanted, although some
light on it might be thrown by the study of the kernels on the disc and the cylinder in
the following sections.

It is apparent from (\ref{fe}) that the number of effective degrees of freedom $M^{2-\frac{\alpha}{2}}$ is the actual order parameter for the cutoff transition. Incidentally, this unusual dependence of the number of degrees of freedom upon $\a$, together with the choice of a large $M$ limit, makes it rather delicate to classify such a transition according to standard terminology.

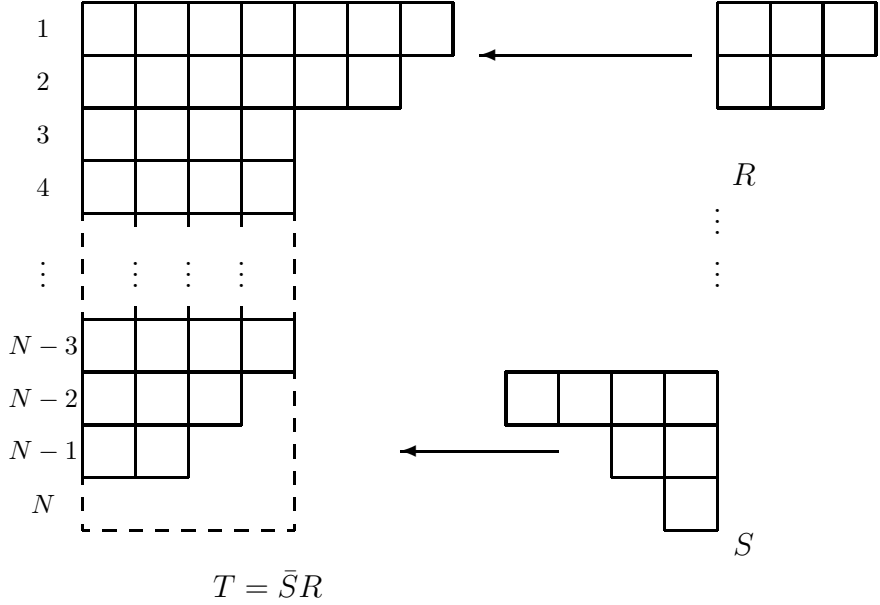
\begin{figure}[tbp]
\centering
\begin{picture}(340,230)(- 190,- 130)
\thicklines

\put(110,35){\makebox(0,0){$R$}}
\put(110,-105){\makebox(0,0){$S$}}
\put(-70,-120){\makebox(0,0){$T = \bar{S}R$}}
\put(-140,-100){\dashbox{5}(80,200)}
\put(90,80){\vector(-1,0){80}}
\put(40,-70){\vector(-1,0){60}}
\put(-155,0){\makebox(0,0){$\vdots$}}
\put(-80,0){\makebox(0,0){$\vdots$}}
\put(-100,0){\makebox(0,0){$\vdots$}}
\put(-120,0){\makebox(0,0){$\vdots$}}
\put(-140,100){\line(1,0){140}}
\put(-140,100){\line(0,-1){80}}
\put(-140,80){\line(1,0){140}}
\put(0,100){\line(0,-1){20}}
\put(-140,60){\line(1,0){120}}
\put(-20,100){\line(0,-1){40}}
\put(-40,100){\line(0,-1){40}}
\put(-140,40){\line(1,0){80}}
\put(-140,20){\line(1,0){80}}
\put(-60,100){\line(0,-1){80}}
\put(-80,100){\line(0,-1){85}}
\put(-100,100){\line(0,-1){85}}
\put(-120,100){\line(0,-1){85}}
\put(-140,-20){\line(1,0){80}}
\put(-140,-20){\line(0,-1){60}}
\put(-140,-40){\line(1,0){80}}
\put(-60,-20){\line(0,-1){20}}
\put(-140,-60){\line(1,0){60}}
\put(-80,-15){\line(0,-1){45}}
\put(-140,-80){\line(1,0){40}}
\put(-100,-15){\line(0,-1){65}}
\put(-120,-15){\line(0,-1){65}}
\put(100,100){\line(1,0){60}}
\put(100,100){\line(0,-1){40}}
\put(100,80){\line(1,0){60}}
\put(160,100){\line(0,-1){20}}
\put(100,60){\line(1,0){40}}
\put(140,100){\line(0,-1){40}}
\put(120,100){\line(0,-1){40}}
\put(100,-40){\line(-1,0){80}}
\put(100,-40){\line(0,-1){60}}
\put(100,-60){\line(-1,0){80}}
\put(100,-40){\line(0,-1){20}}
\put(80,-40){\line(0,-1){20}}
\put(100,-80){\line(-1,0){40}}
\put(80,-60){\line(0,-1){40}}
\put(100,-100){\line(-1,0){20}}
\put(60,-40){\line(0,-1){20}}
\put(60,-60){\line(0,-1){20}}
\put(40,-40){\line(0,-1){20}}
\put(20,-40){\line(0,-1){20}}
\put(100,20){\makebox(0,0){$\vdots$}}
\put(100,0){\makebox(0,0){$\vdots$}}
\put(-155,90){\makebox(0,0){\footnotesize 1}}
\put(-155,70){\makebox(0,0){\footnotesize 2}}
\put(-155,50){\makebox(0,0){\footnotesize 3}}
\put(-155,30){\makebox(0,0){\footnotesize 4}}
\put(-155,-30){\makebox(0,0){\footnotesize $N - 3$}}
\put(-155,-50){\makebox(0,0){\footnotesize $N - 2$}}
\put(-155,-70){\makebox(0,0){\footnotesize $N - 1$}}
\put(-155,-90){\makebox(0,0){\footnotesize $N$}}

\end{picture}
\caption[x]{\footnotesize Young Tableau for a Composite Representation.}
\label{composite}
\end{figure}

Let us finally consider the representations of U($N$) and/or SU($N$) that correspond to
the "momentum" distribution pictured in fig. \ref{prima} and given in (\ref{ro}). 
With the group  SU($N$)\footnote{With SU($N$) the term $\sum_i n_i^2$
in the action should be replaced by $ \sum_i n_i^2 - \frac{1}{N} (\sum_i n_i)^2$ which is
invariant under $n_i \to n_i + a$. However with suitable choice of $a$  the extra term can
be set equal to zero.} this corresponds to a composite representation in the 
sense of ref. ~\cite{Gross:1993yt}, whose Young diagram is shown in fig. \ref{composite}
where the constituent rpresentations are denoted by $R$ and $S$. 
 The rows in the Young diagram of $R$ and $S$ 
 (that coincide in this case) have  lengths  $r_{\alpha}$ and their total number of boxes $|r|$ 
 is given in the large $N$ limit by
 \beq
 |r| =  \sum_{\alpha} r_{\alpha} \sim l^2 \int dx r(x) = \frac{1}{4} l^2
 \label{ytab}
 \eeq
 If the group is U($N$) an arbitrary number of columns of length $N$ can be added or subtracted
 to the Young diagram of fig. \ref{composite}, and the composite representation can be seen as the direct product of the
 two constituent representations $R$ and $S$ with opposite U($1$) charges. In fact the  integers
 labeling the representation $S$ are in this case negative, which corresponds to changing
 signs of the invariant angles $\theta_i$, namely to changing $U$ into $U^{\dagger}$.
 In the strong couplig regime of the Douglas-Kazakov transition ($A>\pi^2$) the partition function is
 dominated in the large $N$ limit by a composite representation of the same type but with the Young
 diagram of  $R$ and $S$ 
 made of rows and columns of lengths of order $N$, rather than $l$, and a total number of boxes of order 
 $N^2$ rather than $l^2$.

\section{Cutoff transition on the disc and on the cylinder}

In this section we are going to consider  the partition function on a disc and on a
cylinder, with fixed holonomies at the boundaries, in the regime where the area $A$ is 
scaled as in (\ref{param}). We shall show that the cutoff transition occurs also in this 
case at the same critical value of $\alpha$, except  for some particular  holonomies at the
boundaries. This is in  analogy to what happens in the case of the Douglas-Kazakov
transition which was proved to occur on a cylinder in \cite{Caselle:1993mq,Gross:1994mr}, 
except possibly for some special configurations (see \cite{Zelditch:2003mt}) .  

We shall use the standard expression for the partition function on a cylinder with
holonomies $U$ and $V$ at the boundaries, which is a particular case of (\ref{first}):
\beq
 Z_{cyl}(U,V)=e^{-\frac{A}{24}(N^2-1)}\sum_{r}\chi_r(U)\chi_r(V)e^{-\frac{A}{2N}\sum_in^2_i}\;.
 \label{parcil}
 \eeq
 The partition function on a disc is obtained from (\ref{parcil}) by taking for instance 
 $U \to 1$ and the partition function on the sphere is recovered by taking the double
 limit $U \to 1$ and $V \to 1$.
 The area $A$ in (\ref{parcil}) scales as in (\ref{param}). The
 sum over the representations $r$ is replaced in the large $N$ limit  by the saddle point, 
 namely by a representation whose "momentum" distribution is of the type described in  fig.
 \ref{prima}. This corresponds to a composite representation, whose constituent representations
 $R$ and $S$ have rows and columns of order $l$ with $l/N \to 0$ in the large $N$ limit.
 The first step in evaluating (\ref{parcil}) is  then to calculate the characters in such
 limit.

\subsection{Characters in the large $N$, large $l$ limit}

For simplicity, let us  consider first the case where only one constituent representation 
is present, which would be the case if only right (or left) moving fermions were present
above the Fermi sea.
This is given by:
\bea
n_i & = & i~~~~~\rm{for}~~~~~ \quad i=1,\ldots, N-l\non
n_i & = & N-l+\alpha+r_\alpha~~~~~{\rm for}~~~~~~~ i=N-l+\alpha~~~~{\rm and}\quad \alpha=1,
\ldots, l\label{chiralrpr}
\eea

The dimension $\Delta_r$ of this representation can be written in the form
\bea
\Delta_r & = & \prod_{i>j}\frac{|(n_i-n_j)|}{(i-j)}=\prod_\a\frac{\a!}{(\a+r_\a)!}
\prod_{\a>\b}\Big(1+\frac{r_\alpha-r_\beta}{\alpha-\beta}\Big)\prod_\a\frac{(N-l+\a+r_\a)!}
{(N-l+\a)!} \non
& = & \frac{d_r}{|r|!}\prod_\a\frac{(N-l+\a+r_\a)!}
{(N-l+\a)!}=  \frac{d_r}{|r|!}N^{\sum r_\a}\Big[1+\mathcal{O}\big(\frac{l}{N}\big)\Big]\;, 
\label{dim2}
\eea
where $d_r$ is the dimension of the representation of the symmetric group $S_{|r|}$ 
associated to the Young diagram  $r$ of rows $r_{\a}$. The order of the symmetric group is
the total number of boxes in the Young diagram: $|r|=\sum_\a r_\a$.  
The dependence on $N$ in (\ref{dim2}) is explicit: the ratio of factorials in (\ref{dim2})
is a polynomial in $N$ of degree $|r|$. If $l$, $\a$ and $r_{\a}$ are all of order $l$ the 
coefficient of  $N^{|r|-k}$  in this polynomial is of order $l^{k}$ in the
large $N$, large $l$ limit .
Hence we can write
\beq
\Delta_r = \frac{d_r}{|r|!}N^{|r|} \Big[ 1 + \sum_k^{|r|} c_k (\frac{l}{N})^k \Big] 
\label{dim3}
\eeq
where the coefficients $c_k$ are smooth in the large $l$, large $N$ limit. If the limit is
taken keeping $l/N$ finite, as in the strong coupling phase of the Douglas-Kazakov
transition, all terms at the r.h.s. of (\ref{dim3}) are of the same order and cannot be
neglegted. 
On the other hand if the double scaling limit is taken with $l/N \to 0$, as in the
previous section, then all terms after the $1$ are subleading and can be neglected.

The same argument holds if instead of the dimension of the representation we consider a
character of  U($N$). In fact the celebrated Frobenius formula gives:
\beq 
\chi_r(U)=\frac{d_r}{|r|!}\sum_{\s\in S_{|r|}}\frac{\hat{\chi}_r(\s)}{d_r} N^{k_{\sigma}}
\prod_j\Big(\frac{\Tr U^{s_j}}{N}\Big)
\label{car}
\eeq
where $\hat{\chi}_r(\s)$ is a character of the symmetric group  $S_{|r|}$ in the
representation labeled by the same Young diagram as $\chi_r(U)$, $s_j$ are the lengths of 
the cycles in the cycle decomposition of $\s$ ($\sum_j s_j=|r|$) and $k_{\s}$ is the 
number of cycles in $\s$. 
By taking $U \to 1$ and comparing (\ref{dim3}) and (\ref{car}) one finds that the
contribution of $\frac{\hat{\chi}_r(\s)}{d_r}$ when $\s$ consists of $k_{\s}$ cycles
is  $ \sim l^{|r|-k_{\s}}$ in the large $l$ limit.
So if we take the double scaling limit where both $l$ and $N$ go to infinity and the
ratio  $l/N$ goes to zero, all terms in the sum over $\s$ in (\ref{car}) are subleading
with respect to the one where $\s$ is the identity. 
 We obtain:
\beq
\chi_r(U)=\frac{d_r}{|r|!} \Big(\frac{\Tr U }{N}\Big)^{|r|} N^{|r|} [1 + {\cal O}(l/N)]=
\Delta_r \Big(\frac{\Tr U }{N}\Big)^{|r|} [1 + {\cal O}(l/N)] 
\label{car2}
\eeq
Notice again that in the strong coupling phase of the Douglas-Kazakov transition, where 
in the large $N$ limit $l/N$ is kept constant, all terms in (\ref{car}) coming from
different permutations $\s$ are of the same order and cannot be neglected.

\subsection{Partition function on the cylinder and special boundary conditions}

As a result of previous analysis we find that in the composite representations
that are dominant in the region $0< \alpha <4$ the characters $\chi_r(U)$ depend only
from $\Tr U$. Hence the partition function on the cylinder will depend only on the trace
of the holonomies on the boundaries. In fact
it is apparent from (\ref{car2}) that going from the sphere ($U$=1) to the cylinder just
amounts to a multiplicative factor $\Big(\frac{\Tr U }{N}\Big)^{|r|}$.
In the case of interest however the composite representation contains both chiral and
anti-chiral  component representations (that is $R$ and 
$S$ of fig. \ref{composite}) and not just one as in the simplified example discussed above. 
However in the large $N$ limit the two component representations are decoupled 
\cite{Gross:1993yt} and the character of the composite representation becomes just the 
product of the characters of the component representations:
\beq
\chi_{composite}(U)= \chi_r(U) \chi_r(U^{\dagger}) = \Delta_r^2  \Big(| \frac{\Tr U }{N} |\Big)^{ 2|r|} 
\label{car3}
\eeq
By replacing (\ref{car3}) into (\ref{parcil}) we find
\beq
Z_{cyl}(U,V) = Z_{sphere} \Big(| \frac{\Tr U }{N} \frac{\Tr V }{N}|\Big)^{ 2|r|}
\label{sphcyl}
\eeq
where of course the value of $|r|$ is the one determined by the saddle point equations and
the equality holds, in the large $N$ limit, only in the regime where the area 
$A$ is scaled as in (\ref{param}) with $\a \neq 0$.
If we set $u=|\frac{\Tr U }{N}|$ and $v=|\frac{\Tr V }{N}|$ it is almost immediate to see
that the multiplicative factor at the r.h.s. of (\ref{sphcyl}) is equivalent to replace
in the action (\ref{action4}) the constant term $\beta$ in the area $A$ with a
$\hat{\beta}(u,v)$ given by:
\beq
\beta \to \hat{\beta}(u,v) = \beta - \log{u^2} - \log{v^2}
\label{betatil}
\eeq
 The partition function on the cylinder is then the same as the partition 
 function on a sphere whose area is 
obtained from the area of the cylinder by adding the two terms $\log{u^2}$ 
and $\log{v^2}$. The latter can be interpreted as the areas of the two discs 
necessary to go, in the given momentum
configuration, from $U=1$ (resp. $V=1$) to the holonomy at the boundary with 
$|\frac{\Tr U }{N}|=u$ (resp $|\frac{\Tr V }{N}|=v$).
The areas of the two discs are of order $1$, so this correction does not affect the
position of the cutoff transition that remains on the cylinder at $\a=4$.

The discussion above relies on the fact that the leading term in  Frobenius formula
(\ref{car}) comes from the identical permutation. 
However this is not always true: if we take the large $N$ and $l$ limit keeping $\Tr U =0$
\footnote{This can happen for instance  if the holonomy at a boundary has a 
symmetry of some sort, for instance of the type $\theta \to \theta+ \pi$, that is
preserved through the limiting process.} then all terms in (\ref{car}) with $\sigma$ 
containing cycles of length $1$ would vanish.
Assuming that $\Tr U^2 \neq 0$, the term in Frobenius formula (\ref{car}) with the highest
 power on $N$  would then come from permutations $\sigma$  {\it all} made out of cycles of
  length $2$ and would be of order $N^{\frac{|r|}{2}}$ instead of $N^{|r|}$. 
  Supposing that both the trace of $U$ and of
 $V$ vanish the coefficient of $\log{M}$ in the first term of (\ref{action3}) would be
 halved. Correspondingly the critical value of $\alpha$ at which the cutoff transition
 occurs would also be halved and become $\alpha_c = 2$
 
 Let us make this argument general and more quantitative. Let us  assume that 
 $\Tr U^{k_1} \neq 0$ with $\Tr U^j=0$  for $j < k_1$ and the same for 
 $V$ with $k_2$ at the place  of $k_1$.
 The leading term in (\ref{car}) will now be:
 \beq
 \chi_r(U) = \frac{d_r}{|r|!} \Big(\frac{\Tr U^{k_1}}{N} \Big)^{\frac{|r|}{k_1}} 
 \frac{\hat{\chi}_r(\s)}{d_r} N^{\frac{|r|}{k_1}}
 \label{asu}
 \eeq
 where $\s$ consists of $\frac{|r|}{k_1}$ cycles of
 length $k_1$\footnote{We assume here for simplicity that we take $l \to \infty$ keeping
 the total number of boxes in the Young diagram multiple of $k_1$ and $k_2$.}
 From the discussion following (\ref{car}) we desume that in the large $l$ limit
 $\frac{\hat{\chi}_r(\s)}{d_r} \sim l^{|r|(1-\frac{1}{k_1})}$
 By using this asymptotic behaviour and eq. (\ref{asu}) we can write the partition
 function on the cylinder in the large $N$ and $l$  limit as in 
 (\ref{action4}), but with $F_0$ and $F_1$ replaced by the following expressions: 
\bea
F_0 & = & (\frac{1}{2 k_1}+\frac{1}{2 k_2}-\frac{\alpha}{4})\int_0^1 dx r(x) \non
F_1 & = & (\frac{1}{2 k_1}+\frac{1}{2 k_2}) \int_0^1 dx r(x) 
\label{fofuno}
\eea
 The dependence of $F_2$ from $r(x)$ is modified, in a so far unknown way, 
 by next to leading terms in  $\frac{\hat{\chi}_r(\s)}{d_r} $\footnote{A lot is known
 on the characters of permutations with cycles all of the same length, so an explicit
 expression for $F_2$ is probably obtainable.} while the constant parameter $\beta$
  is replaced by
 \beq
 \beta \to \hat{\beta}(u_{k_1},v_{k_2}) = \beta - \frac{1}{k_1}\log{u_{k_1}^2} -
  \frac{1}{k_2}\log{v_{k_2}^2}
 \label{betauv}
 \eeq
where, with obvious notations, $u_{k_1}= \frac{|\Tr~U^{k_1}|}{N}$ and $v_{k_2}=
\frac{|\Tr~V^{k_2}|}{N}$.
Although the equation for $r(x)$ cannot be derived, the variation with respect to 
$l$ gives the scaling power of $l$ and the cutoff transition point:
\beq
l \sim M^{1- \frac{\alpha}{4(\frac{1}{2 k_1}+\frac{1}{2 k_2})}}
\label{crcyl}
\eeq
The cutoff transition occurs then at critical point $\alpha_c(k_1,k_2)$ given by:
\beq
\alpha_c(k_1,k_2)= 4(\frac{1}{2 k_1}+\frac{1}{2 k_2})
\label{acry}
\eeq
which generalizes the result on the sphere.

\section{Wilson loops}

In this section we calculate  the expectation value of a Wilson loop in a 
Yang-Mills theory with gauge group U($N$) in the large $N$ limit. 
The space-time manifold has the topology of  a two dimensional sphere whose area
 scales like $\log{N}$ as in eq. (\ref{param}).
 We shall follow th approach of Daul and Kazakov \cite{Daul:1993xz} and 
 Boulatov \cite{Boulatov:1993zs} who did the calculation for constant areas.
 
 We may think of the sphere of area $A$ as two discs of areas $A_1$ and $A_2$ 
 ($A_1+A_2=A$) sewed along their common boundary and  with holonomy on the 
 boundary respectively $U$ and $U^{\dag}$. The Wilson looop is then given by
 
\bea
W(A_1,\,A_2) & = & \langle \frac{1}{N}\Tr U\rangle \non
& = & \frac{1}{Z}\sum_{R_1,R_2}d_1d_2\int dU\frac{1}{N}\Tr U\chi_1(U)\chi_2(U^{\dag})
e^{-\frac{A_1}{2N}C_1-\frac{A_2}{2N}C_2} \label{wloop}
\eea
where $d_1$ and $C_1$ are the dimension and the Casimir operator in the representation 
$R_1$, referred to the disc of area $A_1$; likewise for $d_2$ and $C_2$. The quantity 
$\int dU\Tr U\chi_1(U)\chi_2(U^{\dag})$ may be either $0$ or $1$, namely it is $1$ when 
the Young diagram of $R_2$ is obtained by adding  one box to the diagram of 
$R_1$ and $0$ otherwise. That is, if $R_1$ is labeled by the integers 
$n_1>n_2> \cdots >n_N$,
$R_2$ is labeled by a set on integers where one of the $n_i$ is increased by 
one.
Daul and Kazakov used this property to get rid of one summation and obtained
\bea
W(A_1,\,A_2) & = & \frac{1}{Z}\sum_{R_1}\frac{1}{N}\sum_i d_1^{\,2}\prod_{j,j\ne i}\Big(1+\frac{1}{n_j-n_i}\Big) \non
& & e^{-\frac{A_1+A_2}{2N}C_1}e^{\frac{A_2}{N}n_i}\;, \label{wil}
\eea
where the sum over $i$ corresponds to all the possible ways of adding one box to the 
diagram. This is not however the whole result, in fact the original expression is
symmetric under exchange of $A_1$ and $A_2$, so a term with  $A_1$ and $A_2$
exchanged\footnote{This term originates from the fact that $\int dU\Tr U\chi_1(U)\chi_2
(U^{\dag})$ is different from zero also if the representation conjugate to $R_1$ is 
obtained from the representation conjugate to $R_2$ by adding a box in the Young diagram.} 
must be added to (\ref{wil}) and gives:

\bea
W(A_1,\,A_2) & = & \frac{1}{Z}\sum_{R_1}\frac{1}{N}\sum_i d_1^{\,2}\prod_{j,j\ne i}
\Big(1+\frac{1}{n_j-n_i}\Big) \non
& & e^{-\frac{A_1+A_2}{2N}C_1} \Big(e^{\frac{A_1}{N}n_i}+e^{\frac{A_2}{N}n_i}
\Big)\;. \label{wil2}
\eea
Moreover, eq. (\ref{wil}), as well as (\ref{wil2}), is clearly not symmetric 
under $n_i\leftrightarrow -n_i$, so we can not restrict our considerations to $n_i>0$ 
any longer; instead we have to consider the whole interval $-\infty<n_i<\infty$.

Let us first compute $W(A_1,\,A_2)$ in the \textsl{frozen} phase where the sum 
over $R_1$ is dominated by the trivial representation of dimension $d_1=1$  
labeled  by 
$n_i=i-\frac{N+1}{2}$ $\forall i$ and $R_2$ is the fundamental representation of
 dimension $d_2=N$ with $C_2-C_1=N$. By inserting this into (\ref{wil2}) one 
 finds:
\beq
W(A_1,\,A_2)= \big( e^{-\frac{A_1}{2}} + e^{-\frac{A_2}{2}} \big)
\label{strcou}
\eeq
which is, as expected, a typical strong coupling result. If for instance $A_1 >> A_2$,
then we have
\beq
W(A_1,\,A_2) \sim  e^{-\frac{A_2}{2}}
\label{strcou2}
\eeq
Since $A=A_1+A_2=\a\log N+\b$, this situation can occur in two ways:
\begin{itemize}
\item[1)] $A_2=\b_2$, $A_1=\a\log N+\b_1$: then $W\sim e^{-\frac{\b_2}{2}}$ 
\item[2)] $A_1=\a_1\log N+\b_1$, $A_2=\a_2\log N+\b_2$ with $\a_1>\a_2$: 
then $W \sim  N^{-\frac{\a_2}{2}}$.
\end{itemize}

We proceed now to evaluate $W(A_1,\,A_2)$ in the phase before the cutoff, namely
for $\alpha<4$. The sum over  representations in (\ref{wil2}) can be replaced by
the contribution of the dominant representation in the large $N$ limit,
 calculated on the sphere in section \ref{lA}. The saddle point is unaffected
  by the presence of the extra term   
$$ \sum_i\prod_{j,j\ne i}\Big(1+\frac{1}{n_j-n_i}\Big)\Big(e^{\frac{A_1}{N}n_i}+
e^{\frac{A_2}{N}n_i}\Big) $$
which is  subleading with respect to the action.
After replacing in (\ref{wil2}) the sum with the saddle point contribution, some
 simplifications occur and we get 
\beq \label{wil3}
W(A_1,\,A_2)=\frac{1}{N}\sum_i\prod_{j,j\ne i}\Big(1+\frac{1}{n_j-n_i}\Big)
\Big(e^{\frac{A_1}{N}n_i}+e^{\frac{A_2}{N}n_i}\Big)\;.
\eeq
The representation in (\ref{wil3}) is of the type given in (\ref{rpr}), and the
sum over $i$ describes all possible ways of adding a box to the Young diagram.
However the replacement $n_i \to n_i + 1$   is impossible in the region $-l \leq
i \leq l$ as the resulting sequence of integers would not be monotonic
increasing. So the sum over $i$ in (\ref{wil3}) can be replaced by a sum over 
$\alpha$ with $1 \leq \alpha \leq l$. As a matter of fact we must consider only
positive $\alpha$'s, as adding a box to the adjoint representation amounts to 
symmetrize with respect to $A_1$ and $A_2$, and it has been taken already into account.  
Hence in (\ref{wil3}) we must replace the index $i$ with $\alpha$, $n_i$ with $M-l+\alpha
+ r_{\alpha}$ while the index $j$ goes from $-M$ to $M$, namely it goes over both the
condensed and the non condensed regions.
With these substitutions the expression for the Wilson loop becomes:
\bea
W(A_1,\,A_2) & = & \sum_{\a = 1}^{l} \frac{1}{\xi_{\a}} \big[ 1- \frac{l}{N} 
\big(1-\frac{\xi_{\a}}{l}\big) \big]  \non 
& & e^{\sum_{\beta=1}^{l} \log{\big( 1+ \frac{1}{\xi_{\a} - \xi_{\beta}}\big)} -
\frac{A_2}{2} \big(1 -2 \frac{l-\xi_{\a}}{N}\big)} +\{ A_2 \to A_1\}  
\label{wil4} 
\eea
where 
\beq
\xi_{\a} = \a + r_{\a}
\label{xialpha}
\eeq
It is convenient as usual to use in the large $N$ limit the continuum variables
$x=\frac{\a}{l}$, $\xi(x)= \frac{\xi_{\a}}{l}$ and the density function $\rho(\xi) =
\frac{dx}{d\xi}$.
The crucial part of the calculation is the evaluation of $\sum_{\beta=1}^{l} \log{\big( 1+ \frac{1}{\xi_{\a} 
- xi_{\beta}}\big)}$ which can be done following  ref. \cite{Daul:1993xz}.
By expanding the logarithm one finds:
\beq
\sum_{\beta=1}^{l} \log{\big( 1+ \frac{1}{\xi_{\a} - \xi_{\beta}}\big)}= \int d\eta
\rho(\eta) \frac{1}{\xi - \eta} - \sum_{k=2} \frac{1}{k} \sum_{\beta} \frac{1}{(\beta -
\a)^k } \rho(\xi)^k
\label{sumlog}
\eeq
The first term at the r.h.s. comes from the $k=1$ term of the $\log$ expansion and can be
evaluated using eq. (\ref{rhoeq}), the other terms can be calculated as in 
\cite{Daul:1993xz} and give $\log{\frac{\sin{\pi \rho}}{\pi \rho}}$.
By inserting these results into (\ref{wil4}) and using the explicit form of the solution
(\ref{ro}) one finally obtains (neglecting $O(\frac{l}{N})$ terms):
\beq
W(A_1,\,A_2)= 2 \int d \xi \frac{\sin{\pi \rho}}{\pi} \big( e^{-\frac{A_2}{2}}+
e^{-\frac{A_1}{2}}\big) =  e^{-\frac{A_2}{2}}+ e^{-\frac{A_1}{2}}
\label{wil5}
\eeq
which is exactly the same result as in the frozen phase. The  result is not
trivial, but it was somehow to be expected. Both phases, before and after the
transition, are strong coupling phases and the expectation value of
the Wilson loop should be in both of them obtained, to the leading order,
by filling the loop with elementary plaquettes in the fundamental
representation. The effects of the transition are expected to appear only at the
next-to-leading order (  $\sim \frac{l}{N}$) which is sensitive to the
$O(\frac{l}{N})$ degrees of freedom which are not frozen below the cutoff.

\section*{Acknowledgments}
A.A. wants to thank the "Service de Physique Theorique
(SPhT)" at Saclay for the kind hospitality of these last months. He also would like 
to thank I.~Kostov and G.~Vernizzi for many useful discussions during this period.\\
S.A. would like to thank the Niels Bohr Institute for warm hospitality during the early stage of this work. 

\begin{appendix}
\section{Gauge theories as random walks}

The equivalence between random walks on a group $G$ and two
dimensional gauge theories with gauge group $G$ was pointed out in \cite{D'Adda:2001nf}. 
This equivalence states that the partition function of a gauge theory on a disc of area 
$A$ and holonomy $g$ on the boundary coincides with the probability that a suitably defined 
random walk on the group leads,  in a number of steps proportional to $A$, 
from the identity in the group to the group element $g$. A similar relation 
holds for the partition function on the cylinder.
In ref.  \cite{D'Adda:2001nf} the attention was focused on discrete groups, where the
number of steps can be directly identified with the area measured in suitable units, and 
in particular on the symmetric group $S_n$. In this appendix we extend, in a rather
straightforward way, the argument presented there to the case of gauge theories on
continous Lie groups.
Similar conclusions have in the meantime appeared in the literature \cite{deHaro:2004id}.

Let us consider a random walk on $G$ with the transition probability defined as follows: 
if the walker is in $g_p\in G$ after $p$ steps, then his 
position after $p+1$ steps is obtained by left multiplying $g_p$ by an element 
$g$, chosen in $G$ with a probability $t(g)$.
We assume $t(g)$ to be a class function, whose character expansion can then be written as:

\beq 
t(g)=\sum_{r}\Delta_r  \chi_r(g) \tilde{t}_r\;.
\label{transitionpr}
\eeq 

where $\Delta_r$ is the dimension of the representation $r$.
Suppose that the random walk starts from an element $g_0 \in G$ and denote by $K_p(g,g_0)$
the probability for the walker to be in $g$ after $p$ steps.
Of course $K_0(g,g_0)$ is a delta function, namely:
\beq
K_0(g,g_0) = \sum_r \Delta_r \chi_r(g^{-1}g_0)
\label{zerowalk}
\eeq

Given $K_p(g',g_0)$, the probability for the walker to be in $g$ after $p+1$ steps is given
by:
\beq 
K_{p+1}(g,g_0) =\int dg' ~t\big(g(g')^{-1}\big)K_p(g',g_0)\;,
\label{kp1}
\eeq
where $dg'$ is the Haar measure.
By using (\ref{transitionpr}), (\ref{zerowalk}) and (\ref{kp1}) it is easy to show by induction that
 $K_p(g,g_0)$ is a class function of $g g_0^{-1}$. In fact let us assume that this is true
 for $K_p(g',g_0)$, namely that $K_p(g',g_0)$ admits the character expansion
 \beq
 K_p(g',g_0)= \sum_r \Delta_r k_r^{(p)} \chi_r(g' g_0^{-1})
 \label{kappap}
 \eeq
 Then by replacing (\ref{kappap}) and (\ref{transitionpr}) into (\ref{kp1}) and performing the 
 integration over $g'$ using the characters fusion rules, one finds that $K_{p+1}(g,g_0)$ 
 admits a similar expansion with
 \beq
 k_r^{(p+1)} = \tilde{t}_r  k_r^{(p)}
 \label{kapparec}
 \eeq
 As a result $K_p(g,g_0)$ is given by:
 \beq
 K_p(g,g_0)= \sum_r \Delta_r \chi_r(g g_0^{-1}) \tilde{t}_r^p
 \label{kappap2}
 \eeq
 
 Since we are dealing with a random walk on a continous  Lie group, we expect the walk to be a smooth path. This is obtained by  
 letting the number of steps $p$ go to infinity and at the same time the length of each step to zero. Each step corresponds then to a 
 very small move on the group manifold,  that is $t(g)$ is close to a delta 
 function $\d(g)$, namely according to (\ref{transitionpr}), 
 $\tilde{t}_r\simeq 1$. We implement this  by choosing
\beq
\tilde{t}_r=e^{-\ep h(r)}\,,\quad \textrm{with}\quad \ep\to 0
\label{tsmall}
\eeq

If we  set $g_0 =1$ in (\ref{kappap2}),
\beq
\lim_{\ep\to 0}\lim_{p\to\infty}p\ep = A  \quad \textrm{and} \quad  h(r)=\frac{C_2(r)}{2N}
\label{lmt}
\eeq
we reproduce exactly the partition function of a gauge theory on a disc of area $A$, thus establishing
the desired connection.
A random walk with an arbitrary transition function $h(r)$ will be related to a generalized Yang-Mills theory with the corresponding potential \cite{Douglas:1994pq, Ganor:1994bq} .

The partition function on a cylinder may be associated to the probability of a random walk from a generic
point $g_0$ (in general different from the identity) to $g$. However in order to
 obtain the correct answer we need to consider a random walk
not on the group manifold itself, but rather in the space of orbits, where all group elements belonging to the same equivalent class are
identified. This amounts to replacing in (\ref{kappap2}) $g_0 \to h^{-1} g_0 h$ and integrating over $h$. This produces
a new transition probability $\tilde{K}_p(g,g_0)$ which is separately a class function of both $g$ and $g_0$  and is given by:
\beq
\tilde{K}_p(g,g_0)=  \sum_r \chi_r(g ) \chi_r( g_0^{-1}) \tilde{t}_r^p
\label{kapptilde}
\eeq
This coincides with the kernel on the cylinder, with the parameters identified according to (\ref{lmt}).

\end{appendix}

 \end{document}